\documentclass[aps,pra,twocolumn,showpacs]{revtex4-1}
\usepackage{graphicx}

\usepackage{graphicx,multirow,array,hyperref,amsfonts,epsf}
\begin{document}
\title{Constructing valid density matrices on
an NMR quantum information processor via
maximum likelihood estimation} 
\author{Harpreet Singh}
\email{harpreetsingh@iisermohali.ac.in}
\author{Arvind}
\email{arvind@iisermohali.ac.in}
\author{Kavita Dorai}
\email{kavita@iisermohali.ac.in}
\affiliation{Department of Physical Sciences, Indian
Institute of Science Education \& 
Research (IISER) Mohali, Sector 81 SAS Nagar, 
Manauli PO 140306 Punjab India}
\begin{abstract}
Estimation of quantum states is one of the most
important steps in any quantum information
processing experiment. A naive reconstruction of
the density matrix from experimental measurements
can often give density matrices which are not
positive, and hence not physically acceptable. How
do we ensure that at all stages of reconstruction,
we keep the density matrix positive and
normalized? Recently a method has been suggested
based on maximum likelihood estimation, wherein
the density matrix is guaranteed to be positive
definite. We experimentally implement this
protocol and demonstrate its utility on an NMR
quantum information processor.  We discuss several
examples where we undertake such an estimation and
compare it with the standard method of state
estimation.
\end{abstract}
\keywords{Quantum information processing, Nuclear magnetic resonance,
State reconstruction, Density matrix, Maximum likelihood
method}
\pacs{03.67.Lx}
\maketitle 
\section{Introduction}
\label{intro}
How do we  assign a state to a physical system?  Classically
the answer is very simple: we determine the phase space
point corresponding to the configuration of the system.
This can be achieved by measuring the relevant system parameters 
in an non-invasive manner.  However, for quantum
systems non-invasive measurements are not possible, and
therefore we typically need an ensemble of identically
prepared systems to perform state estimation.  Since the
quantum state cannot be known from a single measurement and
the no-cloning theorem precludes the possibility of making
several copies of the state and using them to make different
measurements on the same state, quantum state estimation is
intrinsically a statistical
process~\cite{massar-prl-95,derka-prl-98}.  Furthermore,
since ensembles are always finite, there is always some
ambiguity associated with the estimated state.
For a given physical situation, we
may even have two different candidate states!  Can these
uncertainties and ambiguities be such that sometimes we end
up estimating the state to be a non-state? Such estimates
should be not allowed and any error or ambiguity in state
estimation should be within the space of positive normalized
density operators. 

Complete estimation of a quantum state from a set of
measurements on a finite ensemble has been a hot topic of
research in quantum information and experimental quantum
computing and several schemes have been proposed and
implemented for quantum state 
estimation~\cite{rehacek-book,plenio-naturecomm,rehacek-pra-15}.  
Standard methods of reconstructing an unknown state
from a set of measurements rely on  quantum state
tomography (QST)
protocols~\cite{helstrom-book,smithey-prl-93,thew-pra-02}.
The QST method  is based on averaging a specialized function
over the experimental data obtained using a quorum of
observables, sufficient to reconstruct the state. However,
the QST
averaging procedure leads to fluctuations which could result
in significant statistical errors as well as an unphysical
density matrix i.e., some eigenvalues could turn out to be negative.
A scheme that redresses this issue of reconstructed
density matrices that are unphysical,
is the maximum-likelihood estimation (MLE)
scheme, which obtains a positive definite estimate for the
density matrix by optimizing a likelihood functional that
links experimental data to the estimated density
matrix~\cite{banaszek-pra-99,hradil-pra-97,kohout-prl-10,plenio-mle}.
The MLE scheme begins with a guess quantum state and
improves the estimate based on the measurements made; the
more the number of measurements, the better is the state
estimate.  A tomographic protocol for two qubits was
recently constructed based on the measurement of 16
generalized Pauli operators which is maximally robust
against errors~\cite{miranowicz-pra-14}.  
A refined iterative
maximum-likelihood algorithm was also proposed to
reconstruct a quantum state and applied to the tomography of
optical states and entangled spin states of trapped
ions~\cite{rehace-pra-07}.  
Other quantum
state estimation algorithms 
include
Bayesian mean estimation~\cite{huszar-pra-12}, least-squares
inversion~\cite{opatrny-pra-97}, numerical strategies for
state estimation~\cite{kaznady-pra-09} and linear regression
estimation~\cite{qi-scirep-13}.

In this work we demonstrate the utility of the MLE scheme
to estimate quantum states on an NMR quantum information
processor.  We experimentally prepare separable and
entangled states of two and three qubits, and reconstruct
the density matrices using both the MLE scheme as well as
QST.  Further, we define an entanglement parameter to
quantify multiqubit entanglement and estimate entanglement
using both the QST and the MLE schemes. We show that while
the QST method overestimates the residual state entanglement
at a given time, the MLE method is able to give us a correct
estimate of the amount of entanglement present in the state.
This is the first demonstration of the advantages of using
MLE for state estimation over QST for NMR quantum
information processing.

The paper is organized as follows: Section~\ref{theory}
contains a concise theoretical description of how to use
QST and MLE
to estimate a quantum state.
Section~\ref{expt} describes the NMR implementation of the
MLE method in estimating quantum states of two and three
qubits and a comparison of results obtained using QST.
Section~\ref{entang} contains a discussion of
using MLE to reconstruct an entangled state on an NMR
quantum information processor. 
Section~\ref{concl} contains some concluding remarks.
\section{Density matrix reconstruction}
\label{theory}
\subsection{Quantum state tomography}
\label{qst}
Quantum state tomography (QST) aims to completely
reconstruct an unknown state via a set of
measurements on an ensemble of identically
prepared states.  Any density matrix $\rho$ of $n$
qubits in a $2^n$-dimensional Hilbert space  can
be uniquely determined using $4^n-1$ independent
measurements and 
the state of the system as described
by its density operator $\rho$ can be reconstructed by performing
a set of projective measurements on multiple
copies of the
state~\cite{chen-pra-13,vandersypen-review}.
Determining all the elements of $\rho$ would involve
making repeated measurements of the same state in different
measurement bases, until all the elements of $\rho$ are
determined~\cite{smithey-prl-93,thew-pra-02,james-pra-01}.

In NMR we cannot perform projective measurements
and instead measure the expectation values of
certain fixed operators over the entire ensemble.
Therefore, we rotate the state via different
unitary transformations before performing the
measurement to collect information about different
elements of the density
matrix~\cite{vandersypen-review}.
The standard tomographic
protocol for NMR  uses the
Pauli basis to expand an $n$ qubit $\rho$,
\begin{equation}
\rho = \sum_{i=0}^{3} \sum_{j=0}^{3} ...
\sum_{k=0}^{3} c_{ij...k} \sigma_i \otimes \sigma_j
\otimes ....\sigma_k
\label{rho-compact}
\end{equation} 
where $c_{00...0}=1$ and $\sigma_0$ denotes the
$2\times 2 $ identity matrix while $\sigma_1$,
$\sigma_2$ and $\sigma_3$ are standard Pauli
matrices. The measurement of expectation values
allowed in an NMR experiment combined with unitary
rotations leads to the determination of the
coefficients $c_{ij..k}$.

In an NMR experiment, we measure the signal
induced in the detection coils while the nuclear
spins precess freely in a strong applied magnetic field.
This signal is called the free induction decay
(FID) and is proportional to the time rate of change of
magnetic flux. This time-domain signal can be expressed in
terms of the expectation values of the transverse
magnetization~\cite{leskowitz-pra-04}:
\begin{equation}
S(t)\propto \mbox{Tr}\left
\{\rho(t)\sum_{k}(I_{kx}-iI_{ky})\right \}
\label{nmrsignal}
\end{equation}
where $I_{kx}=\frac{1}{2^n}(I_1\otimes \cdots
\sigma_{kx}\cdots \otimes I_n) $ and
$\frac{1}{2^n}(I_{ky}=I_1 \otimes \cdots
\sigma_{ky}\cdots \otimes I_n)$ are the Cartesian angular momentum
operators (subset of the product operators) for
the spin $k$ and $\rho(t)$ is the density
operator at time $t$. This signal is then Fourier
transformed to extract information about
expectation values of different operators.  In NMR,
products of a set of the identity operator and the
Pauli spin operators, form the quorum of
observables for
QST~\cite{chuang-proc-98,lee-pla-02,long-joptb}.
While the measured expectation values remain the
same, unitary operators in terms of rf pulses are
applied before measurement to effectively measure
the expectation values of
different operators.

For a system of two qubits, the density matrix
can be expanded in terms of product operators $I_1 \otimes I_2 \dots
\sigma_{1z} \otimes \sigma_{2z}$, as follows:
\begin{eqnarray}
\rho
&=& \frac{1}{4}I_1\otimes
I_2+\langle\sigma_{1x}\otimes I_2\rangle
\sigma_{1x}\otimes I_2+
\dots \nonumber \\
&&+\langle\sigma_{1z}\otimes \sigma_{2y}\rangle
\sigma_{1z}\otimes \sigma_{2y}
+ \langle\sigma_{1z}\otimes \sigma_{2z}\rangle
\sigma_{1z}\otimes \sigma_{2z}
\label{2qrho}
\end{eqnarray}
and $\langle\sigma_{1x}\otimes I_2\rangle$,$\dots$,
$\langle\sigma_{1z}\otimes \sigma_{2z}\rangle$
can be measured by different NMR experiments.
The
FID is collected after the application of 
four unitary operators denoted by 
$\mathbf{I}\mathbf{I}$, $\mathbf{I}X$,
$\mathbf{I}Y$, and $XX$,
where $\mathbf{II}$ corresponds to ``no
operation'' on both spins,
$\mathbf{I}X(Y)$ corresponds to a ``no operation''
on the first spin and a
$90^{\circ}$ rf pulse of phase $X(Y)$ on the 
second spin, and $XX$ corresponds to a
$90^{\circ}$ rf pulse of phase $X$ on both
spins. Fourier transform after this FID collection
leads to the extraction of desired expectation
values.

As an
example for a two-qubit system, we tried creating
the 
the quantum state
$\frac{1}{\sqrt{2}}(\vert 00\rangle
+\vert 01\rangle)$ 
and reconstructed it using standard QST. The 
reconstructed density matrix 
$\rho_{\mbox{QST}}$ turned out to be 
\begin{widetext}
\begin{equation}
{\bf \rho_{\mbox{QST}}}= \left(\begin{array}{cccc}
0.4938&0.5003 +i 0.0014&-0.0221 -i 0.0551&-0.0102 + i0.1282 \\
0.5003 -i 0.0014&0.5062&0.0279 -i 0.1309&0.0168 + i0.0695 \\
-0.0221 + i0.0551&0.0279 + i0.1309&-0.0482&0.0030 - i0.0378 \\
-0.0102 - i0.1282&0.0168 -i 0.0695&0.0030 +i0.0378&0.0482
\end{array}
\right)
\label{qstmatrix}
\end{equation}
\end{widetext}
The above density matrix $\rho_{\mbox {QST}}$ 
reconstructed using
the QST protocol is
normalized and Hermitian  and its
eigenvalues are \{1.0360, 0.0926, -0.0179, -0.1106\}.
As is clear from the last two eigenvalues, the 
reconstructed density matrix is not positive,
and furthermore, $\mbox{Tr}(\rho^2_{\mbox{QST}}) =
1.0944$.  Density matrices that represent physical
quantum states must have the property of positive
definiteness which, in conjunction with the
properties of normalization and Hermiticity,
implies that all the eigenvalues must lie in
the interval [0,1] and  sum to 1 i.e. $0 \leq \mbox{Tr}
(\rho^2 ) \leq 1$. Clearly, the density matrix
reconstructed above by standard QST violates this
condition. Due to its negative eigenvalues it has
as a strange feature that  $\mbox{Tr} (\rho^2) >
\mbox{Tr} (\rho)$.
The obvious reasons for this problem are
experimental inaccuracies, which implies that the
actual magnetization values recorded in an NMR
experiment differ from those that can be obtained
from the Eqn.~(\ref{2qrho}). However, in a correct
estimation scheme the experimental inaccuracies
should lead to an error in the state estimation by
giving a state which is close to the actual state
with some confidence level and should not give a
non-state!  An {\em ad hoc} way to circumvent this
problem is to add a multiple of identity to this
density matrix so that the eigenvalues are
positive. However, this kind of addition is
completely {\em ad-hoc} and leads to non-optimal
estimates and one should be able to do better. We
turn to this issue in the next section via the
maximum likelihood estimation method.
\subsection{Maximum likelihood estimation}
\label{mle}
The example in the previous section
illustrates that most density matrices tomographed using
standard QST may not correspond to a physical
quantum state.
To address this
problem, the
maximum likelihood estimation (MLE) method was
developed to ensure that the reconstructed
density matrix is always positive and
normalized~\cite{james-pra-01}.  The MLE method
estimates the entire quantum state, by finding the
parameters that are most likely to match the
experimentally generated data and maximizing a
specific target function; {\em a priori} knowledge
of the density matrix can also be incorporated
into the method. The main advantage of the method
is that at every stage of the estimation process
the density matrix is positive and normalized and
therefore represents a valid physical situation.

For a system of two qubits, the density matrix can
be written in a compact form following
Eqn.~(\ref{rho-compact}): 
\begin{equation} \rho=
\sum_{j=0}^{3} \sum_{k=0}^{3} n_{jk} \sigma_{j}
\otimes \sigma_{k} \label{rho2} \end{equation}
where $n_{jk}$ are real coefficients determining
the state.

A physical  density matrix $\rho$  has to be
Hermitian, positive  and must have trace equal to
unity. Such a density matrix can  be written
in terms of a lower triangular matrix $T$~\cite{james-pra-01}
\begin{equation}
\rho = T^{\dag} T\quad \mbox{Tr} (T^{\dag} T)=1
\label{tmatrix}
\end{equation}

For the two-qubit system the lower triangular matrix
$T$ from which we obtain $\rho$
has  15
independent real parameters (one parameter from
the 16 is eliminated due to the trace
condition), and can be written in
\begin{equation}
T= \left( \begin{array}{cccc}
t_{1}&0&0&0 \\
t_{5}+it_{6}&t_{2}&0&0 \\
t_{11}+it_{12}&t_{7}+it_{8}&t_{3}&0 \\
t_{15}+it_{16}&t_{13}+it_{14}&t_{9}+it_{10}&t_{4}
\end{array}
\right)
\end{equation}

Given a valid density matrix as described
in~\cite{james-pra-01}, it is possible to 
invert Eqn.~(\ref{tmatrix}) to obtain the matrix $T$ 
\begin{equation}
T= \left( \begin{array}{cccc}
\sqrt{\frac{\Delta}{\mathcal{M}_{11}^{(1)}}}&0&0&0 \\
{\frac{\mathcal{M}_{12}^{(1)}}{\sqrt{\mathcal{M}_{11}^(1)\mathcal{M}_{11,22}^{(2)}}}}&\sqrt{\frac{\mathcal{M}_{11}^{(1)}}{\mathcal{M}_{11,22}^{(2)}}}&0&0
\\
{\frac{\mathcal{M}_{12,23}^{(2)}}{\sqrt{\rho_{44}}\sqrt{\mathcal{M}_{11,23}^{(2)}}}}&\frac{\mathcal{M}_{11,22}^{(2)}}{\sqrt{\rho_{44}}\sqrt{\mathcal{M}_{11,22}^{(2)}}}&\sqrt{\frac{\mathcal{M}_{11,22}^{(2)}}{\rho_{44}}}&0
\\
\frac{\rho_{41}}{\sqrt{\rho_{44}}}&\frac{\rho_{42}}{\sqrt{\rho_{44}}}&\frac{\rho_{43}}{\sqrt{\rho_{44}}}&\sqrt{\rho_{44}}
\end{array}
\right)
\label{rho2t}
\end{equation}
where $\Delta=\mbox{Det}(\rho)$, $\mathcal{M}_{ij}^{(1)}$ is
the first minor of $\rho$ (the determinant of
the $3\times3$ matrix formed by deleting the $i$th and $j$th
columns
of the $\rho$ matrix), $\mathcal{M}_{ij,kl}^{(2)}$ is the second
minor of $\rho$ (the determinant of the $2\times2$
matrix formed by deleting the $i$th and $k$th rows and $j$th and
$l$th columns of the $\rho$ matrix with $i \neq j$ and $k
\neq l$).

From the experimental data we obtain a set of
expectation values $n_{jk}=\langle\sigma_{j}
\otimes \sigma_{k}\rangle= \mbox{Tr}((\sigma_{j}
\otimes \sigma_{k})\rho)$.
It is generally assumed that the experimental
noise has a Gaussian probability distribution and
the probability of obtaining a set of 
measurement results for the set of expectation values
$\{n_{jk}\}$ is \begin{equation}
P(n_{11},\cdots n_{33})= A \prod_{j=0,k=0}^{3,3}
exp\left[-\frac{(n_{jk}-\bar
n_{jk})^2} {2\sigma_{jk}^2}\right]
\end{equation} where $A$ is a normalization
constant and $\sigma_{jk}$ is the standard
deviation of the measured variable $n_{jk}$
(approximately
given by $\sqrt{\bar n_{jk}}$).

The first step in implementing the MLE is the generation of
an initial physically valid density
matrix.  For instance, for a system of two qubits  this
matrix $\rho (t_{1},t_{2},\dots,t_{16})$ is a function
of 16 real variables.  
The next step in the MLE method is to maximize the
likelihood that the physical density matrix $\rho$ will give
rise to the experimental data $\{n_{jk}\}$.  Rather than finding the
maximum value of the probability $P$, the optimization
problem gets simplified by finding the maximum of its
logarithm. Thus the optimization problem is reduced to
finding the minimum of  a ``likelihood function''
\begin{equation}
 \mathcal{L}(t_1,\cdots t_{16})=
\sum_{j=0,k=0}^{3,3}\frac{\left(n_{jk}(t_1,\cdots
t_{16}) - \bar
n_{jk}(t_1,\cdots t_{16})\right)^2}{2\bar\sigma_{jk}^2}
\end{equation}
For a system of two qubits, the optimum set of variables
$\{t_{1}^{opt},t_{2}^{opt},\dots,t_{16}^{opt}\}$ which
minimizes this likelihood function can be determined using
numerical optimization techniques.
We used the Matlab routine ``lsqnonlin''~\cite{matlab} to
find the minimum of the likelihood function. To execute this
routine, one requires the initial estimation of the value of
$t_{1},t_{2},\dots,t_{16}$. Since sixteen 
parameter optimization can be tricky, it is important to
use a good initial guess for parameters. A
reasonable way is to first estimate the state
using the standard method, and obtain the values
of $t_{i}$s using the
Eqn.~(\ref{rho2t}). Since the state may not be
a physically allowed state the parameters obtained
in this manner are not necessarily real. Thus for
our initial guess we drop the imaginary part and
use the real parts of each of $t_{i}$s as the
initial estimate to go into the optimization
routine.  
We used the same 
experimentally generated $\frac{1}{\sqrt{2}}(\vert
00\rangle +\vert 01\rangle)$ state (described in the
example given in the earlier subsection), 
and re-computed the density matrix now
using the MLE
method, and obtained:
\begin{widetext}
\begin{equation}
{\bf \rho_{MLE}}= \left(\begin{array}{cccc}
0.5013 &0.4957 +i0.0011&0.0004 +i 0.0067&0.0003 +i 0.0070 \\
0.4957 -i 0.0011&0.4958&0.0004 +i 0.0067&0.0003 +i 0.0070 \\
0.0004 - i0.0067&0.0004 -i 0.0067&0.0014&0.0015 + i0.0000 \\
0.0003 -i 0.0070&0.0003 -i 0.0070&0.0015 -i0.0000&0.0017
\end{array}
\right)
\label{mlematrix}
\end{equation}
\end{widetext}
The eigen values of this matrix are $\{0.9941,
0.0030, 0.0029, 0.0000\}$ and are all positive and
furthermore $\mbox{Tr}(\rho^2_{\mbox{MLE}})=
0.9883$.  While the density matrix reconstructed
using QST was unphysical, the MLE reconstruction
led to a valid density matrix.
\section{State estimation of two \& three qubits}
\label{expt}
\begin{figure}[h]
\begin{center}
\includegraphics[angle=0,scale=1.0]{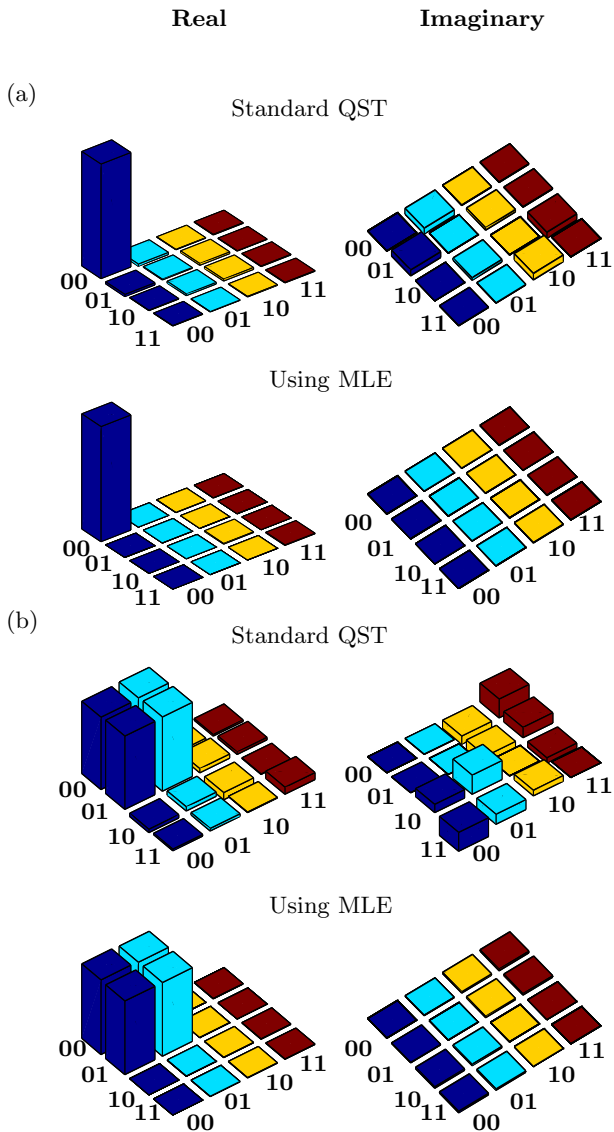}
\caption{ 
Real (left) and imaginary (right) parts of
the experimental tomographs of the (a) $\vert 00 \rangle$
state, with a
computed
fidelity of 0.9833 using standard QST and 
a computed fidelity of 0.9889 using MLE for state estimation.
(b)
$\frac{1}{\sqrt{2}}(\vert 00 \rangle+\vert 01
\rangle)$ state, with a
computed
fidelity of 0.9952 using standard QST and 
a computed fidelity of 0.9939 using MLE for
state estimation.
The rows and columns are labeled in the
computational basis ordered from $\vert 00 \rangle$ to
$\vert 11 \rangle$.}
\label{2tomo}
\end{center}
\end{figure}
\begin{figure}[h]
\centering
\includegraphics[angle=0,scale=1.0]{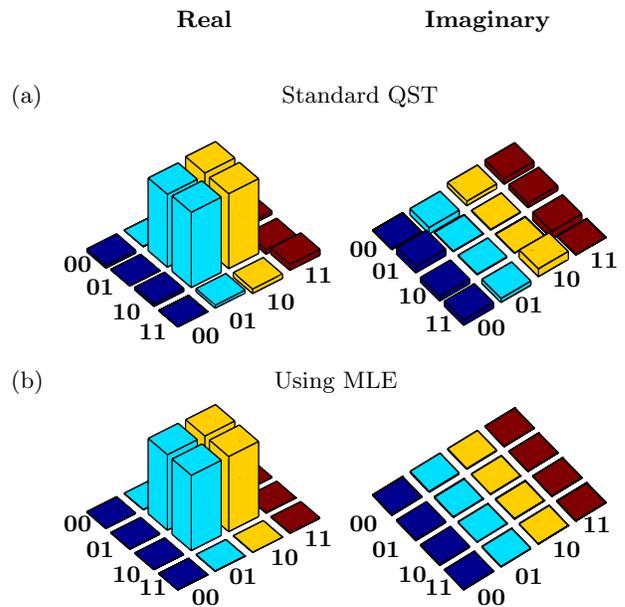}
\caption{
Real (left) and imaginary (right) parts of
the experimental tomographs of the
entangled state $\frac{1}{\sqrt{2}}(\vert 01
\rangle+\vert 10 \rangle)$ reconstructed
(a) using standard QST and (b) using MLE.
The fidelities computed using standard QST
and using MLE for state estimation are
0.9879 
and 0.9976 
respectively. 
The rows and
columns are
labeled in the computational basis ordered from $\vert 00
\rangle$ to
$\vert 11 \rangle$.
}
\label{entangtomo}
\end{figure}
We performed state estimation of several different quantum
states of two and three qubits, constructed on an NMR
quantum information processor, using the MLE method.  The
results were compared every time with the results obtained
by reconstruction using the standard QST protocol. 
The fidelity $F$ of all the states reconstructed
in this section has been computed
using the expression~\cite{weinstein-prl-01}
\begin{equation}
F =
\frac{Tr(\rho_{\rm theory}^{\dag}\rho_{\rm expt})}
{\sqrt(Tr(\rho_{\rm theory}^{\dag}\rho_{\rm theory}))
\sqrt(Tr(\rho_{\rm expt}^{\dag}\rho_{\rm expt}))}
\end{equation}
where $\rho_{\rm theory}$ and $\rho_{\rm expt}$ are the
theoretically expected and experimentally reconstructed
density matrices, respectively.

On a system of two qubits, we began by tomographing a pure
state $|00\rangle $, as well as a superposition state
$\frac{1}{\sqrt{2}}(\vert 00\rangle +\vert 01\rangle)$
(which can be written as a tensor product of the first qubit
in the $\vert 0 \rangle$ state  and the second qubit in a
coherent superposition of the $\vert 0 \rangle$ and $\vert 1
\rangle$ states). The reconstructed density matrices using
the MLE method and using the standard QST method are shown
as bar tomographs in Figure~\ref{2tomo}, with the states
labeled in the computational basis in the order $\vert 00
\rangle$ to $\vert 11 \rangle$.  Using standard QST the
reconstructed $\vert 00 \rangle$ state had negative 
eigenvalues: \{0.9847, 0.0465, 0.0047, -0.0359\} and state fidelity
was computed to be 0.9833. Reconstructing the state using
MLE, we obtained all positive eigenvalues: \{0.9889, 0.0065,
0.0046, 0.0000\}, while state fidelity was computed to be
0.9889.  For the superposition state
$\frac{1}{\sqrt{2}}(\vert 00\rangle +\vert 01\rangle)$,
state reconstruction using standard QST led to some negative
eigenvalues: \{1.036, 0.0926, -0.0179, -0.1106\} with a fidelity
of 0.9952. Using MLE on the other hand, led to all positive
eigenvalues: \{0.9941, 0.0030, 0.0029, 0.0000\} with a state
fidelity  of 0.9939.  While state fidelities are nearly the
same (or slightly better when calculated after MLE
reconstruction of the density matrix), we find that by using
the MLE method for state estimation, we always  obtain a
$\rho$ which is 
physically valid.
\subsection{Estimation of entangled states}
\label{entang}
It has been previously  noted~\cite{james-pra-01} that
the standard QST protocol frequently leads to unphysical
density matrices for entangled multiqubit states. Since
entanglement has been posited to lie at the heart of
quantum computational speedup, their construction and
estimation is of prime importance.  We used the MLE method
to reconstruct two-qubit and three-qubit entangled states
and evaluated the efficacy of this scheme to construct
valid density matrices.

The state estimation of a two-qubit entangled Bell state
$\frac{1}{\sqrt{2}}(\vert 01\rangle+\vert 10\rangle)$  is
shown in Figure~\ref{entangtomo}, using both QST and MLE for
density matrix reconstruction.  Using the QST [protocol for
tomography, we obtain the eigenvalues: \{0.9885, 0.0810,
0.0180, -0.0875\}  with the last
eigenvalue being negative, and with a computed fidelity of 0.9879.  Using
MLE for state estimation leads to all positive eigenvalues:
\{0.9977, 0.0012, 0.006, 0.0005\} with a computed state fidelity
of 0.9976.

\begin{figure}[h]
\centering
\includegraphics[angle=0,scale=1.0]{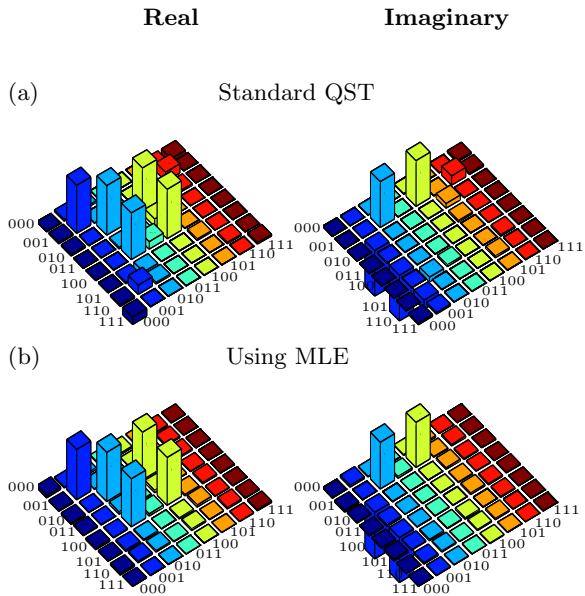}
\caption{ 
Real (left) and imaginary (right) parts of
the experimental tomographs of the three-qubit
maximally entangled state
$\vert W \rangle=\frac{1}{\sqrt{3}}(i\vert
001\rangle+\vert010\rangle+\vert100\rangle)$, 
reconstructed (a) using standard QST
with a
computed
fidelity of 0.9833 and
(b)
using MLE with a
computed
fidelity of 0.9968.
The rows and
columns are labeled in the computational basis ordered from
$\vert 000 \rangle$ to $\vert 111 \rangle$.
}
\label{3tomo}
\end{figure}

Recently, schemes to construct maximally entangled
three-qubit states from a generic state have been
implemented on an NMR quantum information 
processor~\cite{dogra-pra-15,das-pra-15}.
We used these schemes to construct the maximally entangled
$W$ state on a system of three qubits 
$\vert W \rangle=\frac{1}{\sqrt{3}}(i\vert
001\rangle+\vert010\rangle+\vert100\rangle)$, and
thereafter performed state estimation using both
the standard QST and the MLE methods. The experimentally
reconstructed tomographs are depicted in Figure~\ref{3tomo},
with the states being labeled in the computational basis
ordered from $\vert 000 \rangle$ to $\vert 111 \rangle$.
After QST tomography on this three-qubit
state, we obtained the eigenvalues:
\{0.9399, 0.1037, 0.0780, 0.0544, -0.0018, -0.0419, -0.0612, 
-0.0713\}, and a calculated state 
fidelity of 0.9759.   
After performing state estimation using the MLE method, the
eigenvalues turned out to be all positive:
\{0.9191, 0.0361, 0.0267, 0.0075, 0.0064, 0.0024, 0.0015,
0.0000\}, with a calculated state fidelity of 0.9968.

\begin{figure}[t]
\centering
\includegraphics[angle=0,scale=1.0]{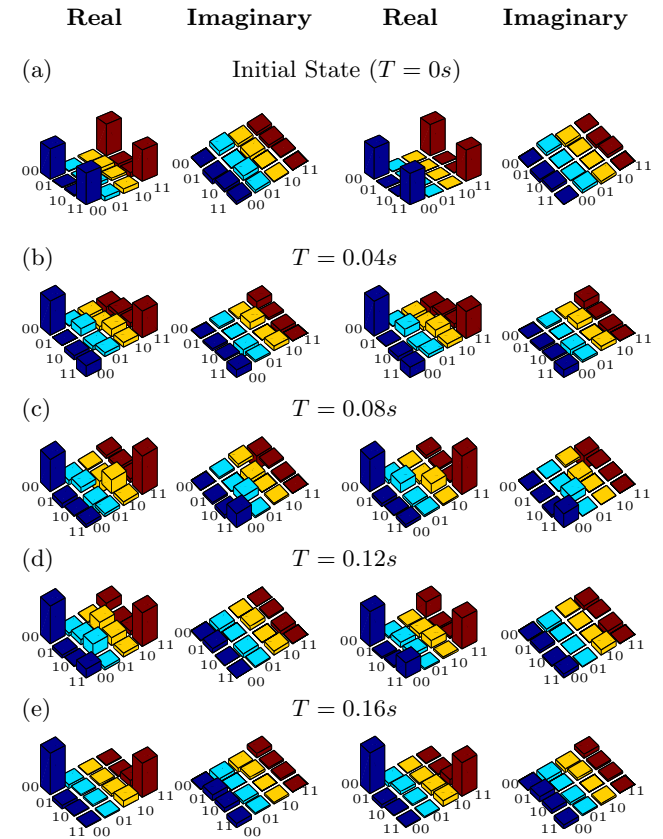}
\caption{ 
Real (left) and imaginary (right) parts of the experimental
tomographs of the (a) 
$\frac{1}{\sqrt{2}}(\vert 00\rangle +\vert 11\rangle)$ state.
(b)-(e) depict the state at $T = 0.04,
0.08, 0.12, 0.16 s$, with the tomographs on the left and the
right representing the state estimated using standard
QST and using
MLE, respectively.  
The rows and
columns are labeled in the computational basis ordered from
$\vert 00 \rangle$ to $\vert 11 \rangle$.
}
\label{fig5}
\end{figure}

A topic of much research focus is the accurate measurement of
the decay of multiqubit entanglement with time. 
To study this, we performed state  estimation
of the entangled two-qubit 
state $\frac{1}{\sqrt{2}}(\vert 00\rangle+\vert 11\rangle)$
using both QST and MLE protocols.
The bar tomographs of the reconstructed density matrices 
at different times ($T=0, 0.04, 0.08, 0.12, 0.16$ sec) 
are shown in Figure~\ref{fig5}.

\begin{figure}[h]
\centering
\includegraphics[angle=0,scale=1.0]{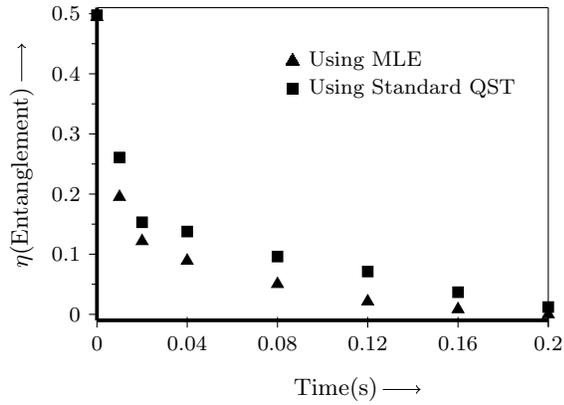}
\caption{Plot of the entanglement parameter $\eta$ with time,
using standard QST and MLE for state reconstruction,
computed for the
$\frac{1}{\sqrt{2}}(\vert 0 0 \rangle + \vert 1 1 \rangle)$
state.}
\label{entangfig}
\end{figure}

The amount of entanglement that remains in the state
after a certain time can be quantified by
an entanglement parameter denoted by $\eta$~\cite{singh-pra-14}.
Since we are dealing with mixed bipartite
states of two qubits, all entangled states
will be negative under partial
transpose (NPT).  
For such NPT states, a reasonable measure of
entanglement is the minimum eigenvalue of 
the partially transposed density operator.
For a given experimentally tomographed 
density operator $\rho$,
we obtain
$\rho^{PT}$ by taking a partial transpose
with respect to one of the qubits.
The entanglement parameter $\eta$ 
for the state $\rho$ in terms of the
smallest eigenvalue $E^{\rho}_{\rm Min}$  
of $\rho^{PT}$ is defined as~\cite{singh-pra-14} 
\begin{equation}
\eta = \left\{\begin{array}{ll}
-E^{\rho}_{\rm Min} &{\rm if~} E^{\rho}_{\rm Min} <0
\\
&\\
\phantom{-}0   &{\rm if~} E^{\rho}_{\rm Min} > 0
\end{array}\right.
\end{equation}
A plot of the entanglement parameter $\eta$ with time
is depicted in Figure~\ref{entangfig}, for the 
two-qubit maximally entangled Bell state
$\frac{1}{\sqrt{2}}(\vert 00\rangle +\vert 11\rangle)$, 
estimated using both standard QST and the MLE method.
As can be seen from Figure~\ref{entangfig}, the QST
method led to negative eigenvalues in the reconstructed
(unphysical) density matrix and hence an overestimation of
the entanglement parameter quantifying the residual
entanglement in the state. The MLE method on the other hand,
by virtue of its leading to a physical density matrix
reconstruction every time, gives us a true measure of
residual entanglement, and hence can be used to
quantitatively study the decoherence of multiqubit
entanglement.
\section{Conclusions}
\label{concl}
We have used the maximum likelihood estimation method
for state estimation on an NMR quantum information
processor, to circumvent the problem of
unphysical density matrices that occur due to
statistical errors while using the standard
QST protocol. 
It has been previously shown that state reconstruction
using QST, of entangled
states and other fragile quantum states are particularly
susceptible to errors, and can lead to unphysical density
matrices for such states.
We show that the experimental density matrices reconstructed for
entangled states of 
two and three qubits using the MLE method are always positive, definite
and normalized. While the state fidelities computed using
QST and using MLE are comparable, the
advantage of the MLE method is that it always leads to a
valid density matrix and hence is a better estimator of the
state of the quantum system.
\acknowledgments
All experiments were performed on a Bruker Avance-III 600
MHz FT-NMR spectrometer at the NMR Research Facility at
IISER Mohali. Arvind acknowledges funding from DST India
under Grant number EMR/2014/000297.
H.S. acknowledges financial support from
CSIR India.
%
\end{document}